# A new cantilever beam-rigid-body MEMS gyroscope: mathematical model and linear dynamics

**Seyed Amir Mousavi Lajimi, Glenn Heppler, Eihab Abdel-Rahman**
Faculty of Engineering, Department of Systems Design Engineering,
University of Waterloo, 200 University Avenue West
Waterloo, Ontario, Canada, N2L 3G1
samousavilajimi@uwaterloo.ca, samirm@sfu.ca

**Abstract -**A new microbeam-rigid-body gyroscope is introduced and its static and dynamic behaviours are studied. The main structure includes a microbeam and an eccentric end-rigid-body influencing the dynamic and static characteristics of the sensor. The sensitivity of the device and the effect of system parameters on the microsystem's response are investigated.

## 1. Introduction

Vibratory gyroscopes work on the basis of the Coriolis effect where the motion in one direction is induced by the motion in the transverse direction modulated by the angular spin rate. Yazdi et al. (1998) provide a comprehensive review of MEMS gyroscopes. Various designs have been proposed to build MEMS gyroscopes. A microbeam gyroscope carrying a point tip-mass was introduced by Esmaeili et al. (2007). By including the electrostatic nonlinearity, Ghommem et al. (2010) completed Esmaeili et al.'s early model. Lajimi and Heppler (2012b; 2012a; 2013) and Lajimi et al. (2009) have studied the effects of end body on the beam-body response.

Herein, a clamped microbeam carrying an end-rigid-body at the other end is considered. The end-rigid-body's significant dimensions affect the system response and require one to account for them. Two electrodes are used to excite and sense the structure's motion in the vertical and horizontal directions, respectively. The excitation signal is composed of a constant and a time-varying voltage, while the sense signal is composed of a constant voltage. We present the mathematical model of the new microsensor and study the statics, linear dynamics, and sensitivity of the device.

## 2. Mathematical Model

The microsystem is composed of a square beam, a rigid microbody attached to beam's end, and two electrodes that apply electrostatic forcing, see Fig. 1. The end body is considered to be rigid with mass $M$ and eccentricity $e$. The eccentricity represents the distance between the beam's end and the center of mass of end-rigid-body. With respect to the local coordinate system axes the end rigid body's rotary inertia matrix includes three diagonal components. According to the extended Hamilton's principle the action integral is

$$I = \int_{t_1}^{t_2} (K - P + W_{nc}) \, dt \qquad (1)$$

where $K$, $P$ and $W_{nc}$ represent respectively the total kinetic energy, the total potential energy and the work of nonconservative forces. The total potential energy including the electrostatic potential energy is given by

$$P = -\frac{1}{2}\int_0^L E\,I\,(\,v''^2 + w''^2)dx$$
$$+2e\,\epsilon\,h\,(\frac{V_v^2}{g_0 - v(L,t) - e\,v'(L,t)} + \frac{V_w^2}{g_0 - v(L,t) - e\,v'(L,t)}) \tag{2}$$

where the proposed form of the electrostatic potential energy is computed at the end-rigid body's center of mass. In Eq. (2), $v(x,t)$ and $w(x,t)$ represent the flexural displacements and $V_v$ and $V_w$ the actuation voltages in the $y$ and $z$ directions, respectively. Initial gap size, the elastic modulus, and the second moment of area are respectively indicated using $g_0$, $E$, and $I$. The total kinetic energy is expressed as

$$K = K_B + K_M$$
$$= \frac{1}{2}M\dot{\boldsymbol{R}}\cdot\dot{\boldsymbol{R}} + \dot{\boldsymbol{R}}\cdot\left(\boldsymbol{\omega}(L,t)\times\int_M \boldsymbol{\rho}\;\mathrm{d}M\right)$$
$$+ \frac{1}{2}\boldsymbol{\omega}(L,t)\cdot\int_M \boldsymbol{\rho}\times(\boldsymbol{\omega}(L,t)\times\boldsymbol{\rho})\;\mathrm{d}M + \int_0^L m\,\dot{\boldsymbol{r}}_p\cdot\dot{\boldsymbol{r}}_p\;\mathrm{d}x \tag{3}$$

where $M$ is the total mass of the end body, $\dot{\boldsymbol{R}}$ the velocity vector, computed relative to the base frame, of a reference point in the end rigid body chosen to coincide with end of the beam, $\boldsymbol{\omega}(L,t)$ is the angular velocity of the beam's section at $L$, $\boldsymbol{\rho}$ is the position vector, relative to the end of the beam, of an arbitrary point in the end rigid body, $m$ is the mass per unit length of the beam, $p$ is an arbitrary point on the beam's cross-section, and $\dot{\boldsymbol{r}}_p$ is the velocity vector for the arbitrary point $p$ relative to the inertial base frame, see Fig.1, $\times$ denotes the vector cross product, and $\cdot$ the inner product.

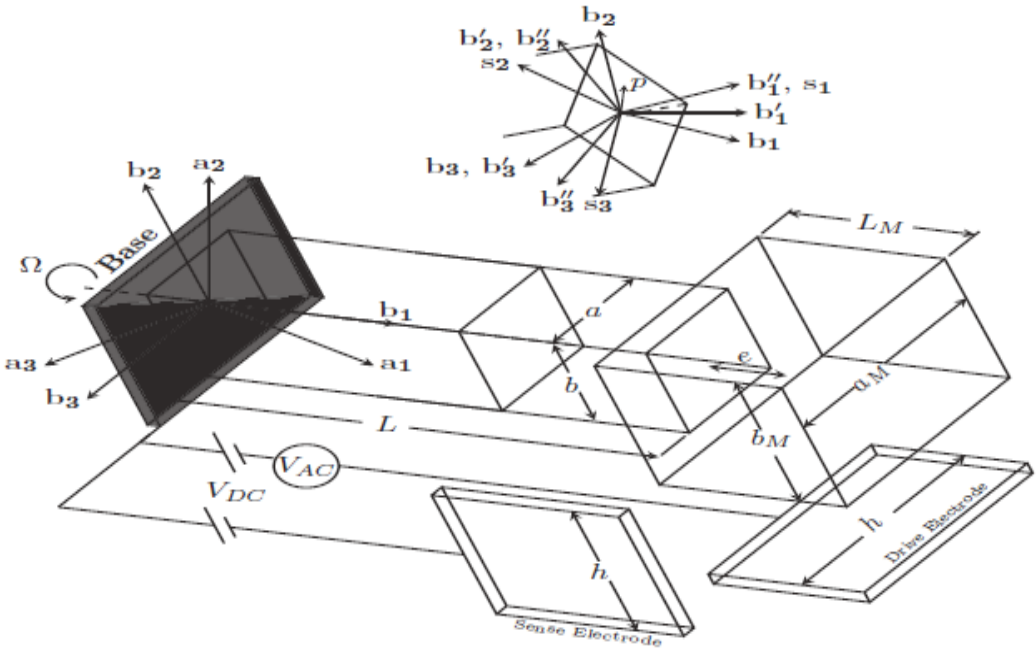

Fig. 1 The microgyroscope: the gyroscope rotates about the longitudinal axis. Systems of inertial $(a_1, a_2, a_3)$, base $(b_1, b_2, b_3)$, and sectional $(s_1, s_2, s_3)$ frames are used to obtain the mathematical model of MEMS gyroscope. The eccentricity (the distance between beam's end and end-rigid-body's center of mass) is denoted by $e$.

Table 1 System parameters

| $L$ $(\mu m)$ | $a$ $(\mu m)$ | $b$ $(\mu m)$ | $L_M$ $(\mu m)$ | $a_M$ $(\mu m)$ | $b_M$ $(\mu m)$ | $h$ $(\mu m)$ | $g$ $(\mu m)$ | $e$ $(\mu m)$ | $E$ (GPa) | $\rho$ $(\frac{kg}{m^3})$ | $\epsilon$ $(\frac{F}{m})$ |
|---|---|---|---|---|---|---|---|---|---|---|---|
| 300 | 5 | 5 | 50 | 20 | 5 | 5 | 2 | 25 | 160 | 2330 | $8.854$ $\times 10^{-12}$ |

Substituting Eqs. (2) and (3) into (1) including the proportional Raleigh damping term, and setting the variation of the functional to zero, results in the following equations of motion in the sense direction

$$EI\,v''' + m\,\ddot{v} - 2m\,\Omega\,\dot{w} - m\,\dot{\Omega}\,w - m\Omega^2 v - J\Omega^2 v'' - J\dot{\Omega}w'' - J\ddot{v}'' = 0 \quad (4)$$

The boundary conditions at $x=0$ are

$$v(0) = 0, \quad v'(0) = 0 \quad (5)$$

and, at $x=L$ are

$$EI\,v''' + M\big(\Omega^2 v + \dot{\Omega}\,w + 2\,\Omega\,\dot{w} - \ddot{v} + e\,\Omega^2 v' + 2e\,\Omega\,\dot{w}' - e\,\ddot{v}'\big) - J_{22}\dot{\Omega}v' \quad (6)$$

$$- J_{33}\Omega^2 v' - J_{33}\ddot{v}' = -\frac{\epsilon\,A_v V_v^2}{2(g_v - v - e\,v')^2}$$

$$EI\,v'' - M\big(e\,\Omega^2 v + e\,\dot{\Omega}\,w + 2e\,\Omega\,\dot{w} - e\,\ddot{v} + e^2\,\Omega^2 v' + e^2\dot{\Omega}w' + 2\,e^2\,\Omega\,\dot{w}' \quad (7)$$

$$- e^2\,\ddot{v}'\big) - J_{22}(\Omega^2 v' + \Omega\dot{w}') + J_{11}\big(\Omega^2 v' + \dot{\Omega}\,w' + \Omega\,\dot{w}'\big)$$

$$- J_{33}\big(\dot{\Omega}w' + \Omega\dot{w}' - \dot{v}''\big) = \frac{\epsilon\,e\,A_v V_v^2}{2(g_v - v - e\,v')^2}$$

where the electrostatic forcing areas are denoted by $A_v$ in the sense direction. The gap distance between the side electrode and the end body's outer surface at zero bias voltage is $g_v$, the components of the rotary inertia of end-rigid-body relative to its center of mass are $J_{11}$, $J_{22}$ and $J_{33}$, respectively. Similar equations are found for the drive direction by replacing $v$ with $w$ and $w$ with $v$ in Eqs. (4)-(7). The boundary condition equations are for the general case of a cantilever beam rigid-body gyroscopic system and can be simplified to the beam-tip mass case by setting $e$, $J_{11}$, $J_{22}$, and $J_{33}$ all to zero.

## 3. Static analysis

To obtain the static equations the time derivatives in Eqns. (4)-(7) are set to zero. Similar equations are obtained for both $w$ and $v$. Thus, the following static equation

$$EI\,v''' - m\Omega^2 v - J\Omega^2 v'' = 0 \quad (8)$$

subject to

$$v(0) = 0, \quad v'(0) = 0 \quad (9)$$

and

$$EI\,v''' + M\big(\Omega^2 v + e\,\Omega^2 v'\big) - J_{33}\Omega^2 v' = -\frac{\epsilon\,A_v V_v^2}{2(g_v - v - e\,v')^2} \quad (10)$$

$$EI\,v'' - M(e\,\Omega^2 v + e^2\,\Omega^2 v') - J_{22}\Omega^2 v' + J_{11}\Omega^2 v' = \frac{\epsilon\,e\,A_v V_v^2}{2(g_v - v - e\,v')^2} \quad (11)$$

is obtained. System parameters are presented in Table 1. In Fig. 2 the static displacement at the tip of the end rigid body is plotted versus the DC voltage. It can be seen that angular spin rate does not have a significant effect in sifting the curve to the left. Thus, the pull-in voltage almost remains constant. Solid (blue) line indicates the stable static equilibrium position while dashed (red) line the unstable position.

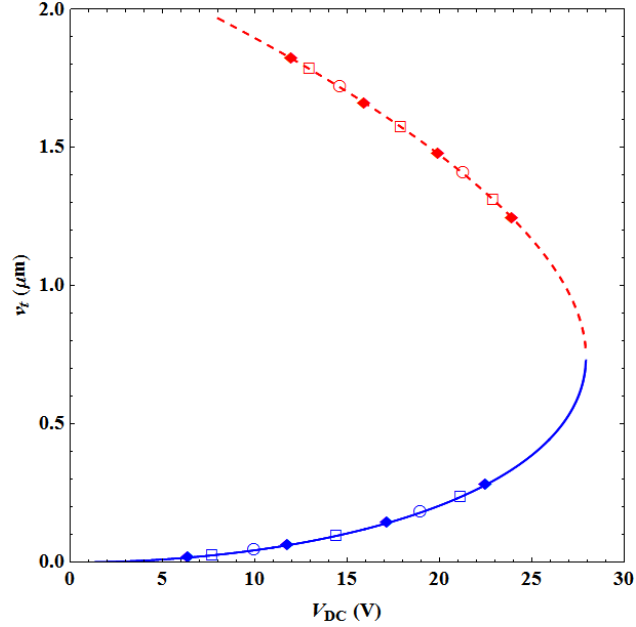

Fig. 2 Static displacement at the tip of the end body vs. the angular spin rate: -♦- for $\Omega = 0$ Hz, -○- for $\Omega = 100$ Hz, and -□- for $\Omega = 1000$ Hz, unstable branch is denoted by dashed line

Although the current form of electrostatic moment and force in Eqns. (6) and (7), respectively, are an approximation of distributed electrostatic force on the corresponding surface of end-rigid-body, they give an accurate estimation of the stable branch of static displacement (solid blue line in Fig. 2). To compute the exact electrostatic force and moment, distributed force and moment are integrated on the active area and hence moment and shear force boundary conditions are reformed

$$EI\, v''' + M(\Omega^2 v + e\, \Omega^2 v') - J_{33}\Omega^2 v' = -\frac{\epsilon\, e\, h\, V_v^2}{(g_v - v - 2\, e\, v')(g_v - v)} \tag{12}$$

$$\begin{aligned} EI\, v'' - M(e\, \Omega^2 v + e^2\ \Omega^2 v') - J_{22}\Omega^2 v' + J_{11}\Omega^2 v' \\ = \frac{\epsilon\, h\, V_v^2}{2\, v'^2}\left( \mathrm{Ln}\left(\frac{g_v - v - 2\, e\, v'}{g_v - v}\right) + \frac{2\, e\, v'}{g_v - v - 2\, e\, v'}\right) \end{aligned} \tag{13}$$

Computing the static response using Eqns. (12) and (13) instead of Eqns. (10) and (11) shows that Eqns. (10) and (11) provide an excellent approximation of stable static response. While using Eqns. (12) and (13) is not required for practical purposes, if one needs to obtain an accurate estimate of unstable static equilibrium Eqns. (12) and (13) should be used requiring more powerful computational resources.

## 4. Dynamic analysis

To perform a dynamic analysis, the response is separated into static and dynamic components as

$$w(x,t) = w_s(x) + \psi(x)q(t) \text{ and } v(x,t) = v_s(x) + \phi(x)p(t) \tag{14}$$

Substituting Eq. (8) into the Lagrangian of the system and using Lagrange's equations of motion, adding the proportional damping terms to each equation, and linearizing the electrostatic force about the static equilibrium positions the dynamic equations are obtained. Mode shapes $\psi(x)$ and $\phi(x)$ are obtained by setting the end mass parameters to zero in the equation of motion and solving the frequency equation for a simple cantilever beam.

One of the most important parameters in the design of MEMS and particularly MEMS sensors is the quality factor which is defined as $Q = \omega_n\, m/c$ where $\omega_n$ is the natural frequency of the system, $m$ the mass, and $c$ is the damping factor. On the other hand, a critical performance indicator is the sensitivity of the device. Sensitivity is directly proportional to the ratio of dynamic response in the sense direction to the dynamic response in the drive direction. In Fig. 3(a) and 3(b) the ratio of drive vibration amplitude to the sense vibration amplitude are plotted for low and moderately high quality factors for a range of excitation frequency and angular rotation rate. The higher the quality factor the sharper the resonance curve is. The input angular rotation rate $\Omega$ varies between 0 and 36000 °/sec.

The normalized excitation frequency $\widehat{\Omega}_e$ is defined as the ratio of the excitation frequency to the natural frequency of the beam. The excitation frequency is set to match natural frequency in the sense direction. Equal bias DC voltages are applied in both drive and sense directions causing small natural frequency mismatch between natural frequencies in the drive and sense directions. However, the beam and end-rigid-body are designed such that natural frequencies do not differ more than one Hz from each other at zero DC voltage. Although matching natural frequencies improves the sensitivity of MEMS gyroscopes, other performance requirements may prevent one from exact matching of natural frequencies (2006).

## 4. Conclusion

We have presented a mathematical model of a new cantilever beam-rigid-body microgyroscope. The static and dynamic behaviour of the new MEMS gyroscope has been studied. It has been shown that the sensitivity and performance of the device is influenced by the quality factor (the damping ratio). Therefore, the operating range of the sensor should be chosen based on the application.

## References

Apostolyuk, V. (2006). Theory and design of micromechanical vibratory gyroscopes. Mems/nems, Springer**:** 173-195.

Esmaeili, M., Jalili, N. and Durali, M. (2007). "Dynamic modeling and performance evaluation of a vibrating beam microgyroscope under general support motion." Journal of Sound and Vibration **301**(1-2): 146-164.

Ghommem, M., Nayfeh, A. H., Choura, S., Najar, F. and Abdel-Rahman, E. M. (2010). "Modeling and performance study of a beam microgyroscope." Journal of Sound and Vibration **329**(23): 4970-4979.

Lajimi, A. M., Abdel-Rahman, E. and Heppler, G. R. (2009). On natural frequencies and mode shapes of microbeams. in Proceedings of the International MultiConference of Engineers and Computer Scientists, March 18 - 20, Hong Kong.

Lajimi, S. A. M. and Heppler, G. R. (2012a). "Comments on "natural frequencies of a uniform cantilever with a tip mass slender in the axial direction"." Journal of Sound and Vibration **331**(12): 2964-2968.

Lajimi, S. A. M. and Heppler, G. R. (2012b). Eigenvalues of an axially loaded cantilever beam with an eccentric end rigid body. in International Conference on Mechanical Engineering and Mechatronics, August 16-18, Ottawa, Canada**:** 135-131-135-138.

Lajimi, S. A. M. and Heppler, G. R. (2013). "Free vibration and buckling of cantilever beams under linearly varying axial load carrying an eccentric end rigid body." Transactions of the Canadian Society for Mechanical Engineering **Accepted for publication**.

Yazdi, N., Ayazi, F. and Najafi, K. (1998). "Micromachined inertial sensors." Proceedings of the Ieee **86**(8): 1640-1658.

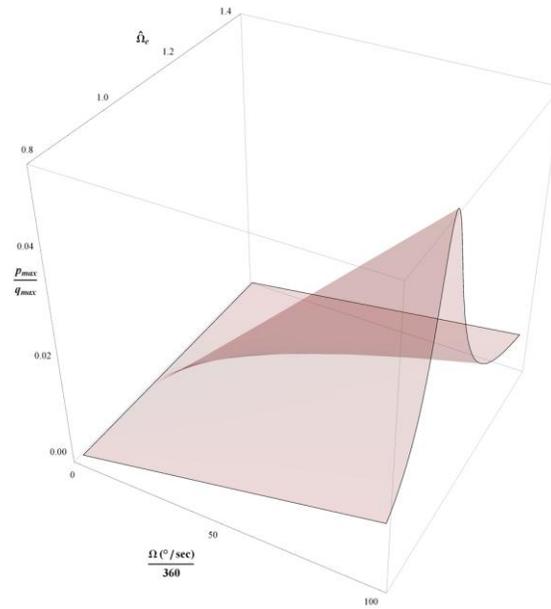

(a)

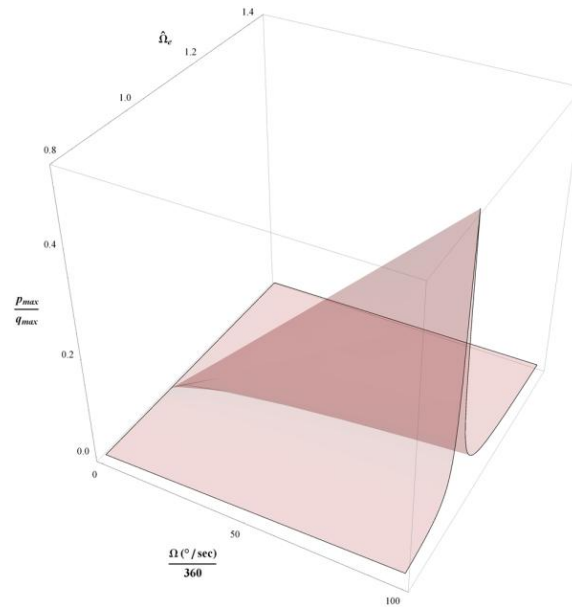

(b)

Fig. 3 Sensitivity vs. the angular rotation rate and the normalized excitation frequency for $V_{DC} = 10$ V in both drive and sense directions and $V_{AC} = 3$ V. (a) Q = 20 and (b) Q = 200